\documentstyle[12pt]{article}
\begin{document}
\parskip 10pt plus 1pt
\title{Duality Invariance of Cosmological Solutions with Torsion}
\author{
{\it  Debashis Gangopadhyay}\\
{S.N.Bose National Centre For Basic Sciences}\\
{JD Block, Sector-III, Salt Lake, Calcutta-700091, INDIA}\\
{\it Soumitra Sengupta}\\
{Department of Physics, Jadavpur University}\\
{Calcutta-700032, INDIA}\\
}
\date{}
\maketitle
\baselineskip=20pt

\begin{abstract}
We show that for a string moving in a background consisting of
maximally symmetric gravity, dilaton field and second rank
antisymmetric tensor field, the $O(d) \otimes O(d)$
transformation on the vacuum solutions gives inequivalent
solutions that are not maximally symmetric. We then show that the usual
physical meaning of maximal symmetry can be made to remain unaltered
even if torsion is present and illustrate this through two toy
models by determining the torsion fields, the metric and Killing vectors.
Finally we show that under the $O(d) \otimes O(d)$ transformation this
generalised maximal symmetry can be preserved under certain conditions.
This is interesting in the context of string related cosmological backgrounds.
\end{abstract}
\newpage

\centerline {\bf I.Introduction}

Some time back it was shown that the low energy string effective
action possesses, for time dependent metric $G_{\mu\nu}$, torsion
$B_{\mu\nu}$  and dilaton $\Phi$ background fields $(\mu, \nu =
1, 2, ..d)$ a full continuous O(d,d) symmetry ( a genaralisation of
T-duality in string theory) under which
"cosmological" solutions of the equations of motion are
transformed into other inequivalent solutions $^{1}$. Subsequently,
a generalisation to this was obtained $^{2}$. These transformations
are conjectured to be a generalisation of the Narain construction
$^{3}$ to curved backgrounds.

Here we investigate the consequences of this O(d) $\otimes$ O(d)
transformation on the space-time symmetries of the theory. We
consider a string propagating in a gravity, dilaton and second
rank antisymmetric tensor background and show that if the full
metric corresponding to a given background is maximally symmetric
then under the O(d) $\otimes$ O(d) twist this symmetry is not
preserved. However, in a generalised definition of maximal
symmetry when torsion is present, we show that this generalised
maximal symmetry can be preserved under certain conditions.

We first discuss the meaning of maximal symmetry and the
O(d) $\otimes$ O(d) symmetry. We then show that an approximate
maximally symmetric solution with $B_{\mu\nu} \not=0$ and non-zero curvature
is possible with a linear dilaton background. However, this symmetry is
destroyed under O(d)$\otimes$ O(d) twist. Finally, we give a generalisation
to the meaning of maximal symmetry when torsion is present and show that
this generalised maximal symmetry can be preserved under the O(d)
$\otimes$ O(d) twist if the torsion fields satisfy certain conditions.

For a maximally symmetric space-time $^{4}$ the curvature
$$R_{iklm} = K\left(g_{im}\enskip g_{kl} - g_{il}\enskip g_{km}\right)
\eqno(1)$$
K is the curvature constant proportional to the scalar curvature
$R^{i}_{i}$. Two maximally symmetric metrics with the same K and
the same number of eigenvalues of each sign, are related by a
coordinate diffeomorphism $^{4}$.

\centerline {\bf II.The O(d) $\otimes$ O(d) Invariance of the String Effective Action}

Now consider the low energy effective action of string theory in
D space-time dimensions.
This is $^{5}$ :
$$S = - \int d^{D}X \sqrt{det G} e^{-\phi} \left[\Lambda -
R^{(D)} (G) + \left({1 \over 12}\right) H_{\mu\nu\rho}H^{\mu\nu\rho} -
G^{\mu\nu} \partial_{\mu}\Phi\partial_{\nu}\Phi \right] \eqno(2)$$
where $H_{\mu\nu\lambda} = \partial_{\mu} B_{\nu\lambda} +$
cyclic perm.$R^{(D)}$ is the D-dimensional Ricci scalar and
$\Lambda$ is the cosmological constant equal to $\frac{(D -
26)}{3}$ for the bosonic string and $\frac{(D - 10)}{2}$ for the
fermionic string. It has been shown that $^{2}$ for (a) $X \equiv
\left(\hat{Y}^{m}, \tilde{Y}^{\alpha}\right), 1 \leq m \leq d, 1 \leq
\alpha \leq D-d$. $\hat{Y}^{m}$ having Euclidean signature,
(b) background fields independent of $\hat{Y}^{m}$, and \\
(c) $G = \left(\matrix{
\hat{G}_{mn}	&0\cr
0		&\tilde{G}_{\alpha\beta}
\cr}\right) \enskip ; \enskip
B = \left(\matrix{
\hat{B}_{mn}	&0\cr
0		&\tilde{B}_{\alpha\beta}
\cr}\right)$
the action (1) can be recast into
$$S = - \int d^{D}\hat{Y}\int d^{D-d}\tilde{Y}\sqrt{det}\tilde{G}e^{-\chi}
\phantom{..........................}$$
$$[\Lambda -
\tilde{G}^{\alpha\beta}\tilde{\partial}_{\alpha}\chi
\tilde{\partial}_{\beta}\chi - \left({1 \over 8}\right)
\tilde{G}^{\alpha\beta} Tr \left(\tilde{\partial}_{\alpha} ML
\tilde{\partial}_{\beta} ML\right)
-\tilde{R}^{(D-d)}(G) + \left({1\over12}\right)\tilde{H}_{\alpha\beta\gamma}
\tilde{H}^{\alpha\beta\gamma}] \eqno(3)$$
where
$$L = \left(\matrix{
0	&1\cr
1	&0
\cr}\right) \eqno(4)$$
$$\chi = \Phi - ln \sqrt{det \hat{G}} \eqno(5)$$
$$M = \left(\matrix{
\hat{G}^{-1}		&- \hat{G}^{-1} \hat{B} \cr
\hat{B} \hat{G}^{-1}	&\hat{G} - \hat{B} \hat{G}^{-1} \hat{B}
\cr}\right) \eqno(6)$$

If one of the coordinates $\hat{Y}^{1}$ is time-like, then the
action $(2)$ is invariant under an $O(d-1, 1) \otimes O(d-1, 1)$
transformation on $\hat{G}, \hat{B}$ and $\Phi$ given by
$$M \rightarrow \left({1 \over 4}\right) \left(\matrix{
\eta(S+R)\eta	&\eta(R-S)\cr
(R-S)\eta	&(S+R)
\cr}\right)
M
\left(\matrix{
\eta\left(S^{T}+R^{T}\right)\eta	&\eta\left(R^{T}-S^{T}\right)\cr
\left(R^{T}-S^{T}\right)\eta		&\left(S^{T}+R^{T}\right)
\cr}\right) \eqno(7)$$
with $\eta$ = diag (-1, 1, ........1) ; S, R some O(d-1, 1)
matrices satisfying $S \eta S^{T} = \eta$,$ R \eta R^{T} = \eta$ ;
$S_{11} = cosh \theta = R_{11}$ ,$ S_{21} = -sinh \theta = -R_{21}$,
and $  S_{i1} = R_{i1} = 0 $,	for $ i \geq 3$.

In component form the transformed fields are given by $^{2}$ :
$$\left(\hat{G}^{'-1}\right)_{ij} = \left({1 \over
4}\right)[\eta(S+R)\eta \hat{G}^{-1}\eta\left(S^{T}+R^{T}\right)\eta$$
$$ + \eta(R-S) \left(\hat{G} -
\hat{B}\hat{G}^{-1}\hat{B}\right)\left(R^{T}-S^{T}\right)\eta$$
$$ -\eta(S+R)\eta \hat{G}^{-1}\hat{B} \left(R^{T}-S^{T}\right)\eta$$
$$ +\eta(R-S) \hat{B}\hat{G}^{-1}
\eta\left(S^{T}+R^{T}\right)\eta]_{ij} \eqno(8a)$$
$$\left(\hat{B'}\right)_{ij} = \left({1 \over 4}\right)[\{(R-S)\eta
\hat{G}^{-1}\eta \left(S^{T}+R^{T}\right)\eta$$
$$ + (S+R)\left(\hat{G} - \hat{B}\hat{G}^{-1}\hat{B}\right)
\left(R^{T}-S^{T}\right)\eta$$
$$ + (S+R) \hat{B} \hat{G}^{-1}\eta
\left(S^{T}+R^{T}\right)\eta$$
$$ - (R-S)\eta \hat{G}^{-1} \hat{B}
\left(R^{T}-S^{T}\right)\eta\}\hat{G'}]_{ij} \eqno(8b)$$
$$\Phi' = \Phi - \left({1\over 2}\right) ln\enskip\ det \hat{G} +
\left({1\over 2}\right) ln\enskip\ det \hat{G'} \eqno(8c)$$
The equations of motion obtained from $(2)$ are
$$R_{\mu\nu} = D_{\mu}D_{\nu}\Phi + \left({1\over 4}\right)
H^{\lambda\rho}_{\mu} H_{\nu\lambda\rho} \eqno(9a)$$
$$D_{\mu}\Phi D^{\mu}\Phi - 2 D_{\mu}D^{\mu}\Phi + R -
\left({1\over 12}\right) H_{\mu\nu\rho} H^{\mu\nu\rho} = 0 \eqno(9b)$$
$$D_{\lambda} H^{\lambda}_{\mu\nu} - \left(D_{\lambda}
\Phi\right) H^{\lambda}_{\mu\nu} = 0 \eqno(9c)$$

\centerline {\bf III.Construction of Maximally Symmetric Solutions}

Maximal symmetry implies
$$R_{\mu\nu} = K(1 - D) g_{\mu\nu} \eqno(10)$$
i.e.\enskip\ $$R = K(1 -D)D \eqno(11)$$

We can show that for $B_{\mu\nu} = 0$, the only maximally
symmetric solution is that for which the dilaton background is a
constant and the curvature constant $K = 0$ $^{6}$. Here we shall
discuss in detail the more general case where $B_{\mu\nu} \not= 0$.

For $B_{\mu\nu} \not= 0$, maximally symmetric solutions to $(9)$
are for $D = 3$	 $^{6}$
$$H^{2}_{01r} = \frac{K(1 - D)}{2} g_{00}\enskip\ g_{11} \eqno(12a)$$
i.e.$$ B_{01} = \left[\frac{K(1 - D)}{2}\right]^{1\over 2} \int
dr \left[g_{00}\enskip\ g_{11}\right]^{1\over 2} \eqno(12b)$$
and$$ \partial_{r}^{\enskip\ 2} \Phi = 0\enskip\ i.e.\enskip \Phi =
\alpha\enskip\ r + \beta \eqno(12c)$$

Now consider the case of $B_{\mu\nu} \not= 0$. We take the
metric as
$$ds^{2} = - f_{0}(r) dt^{2} + f_{1}(r) \left(dx_{1}\right)^{2} + dr^{2} \eqno(13a)$$
$$f_{0}(r) = cos^{2}\left(\sqrt{K_{1} r}\right), f_{1}(r) =
sin^{2}\left(\sqrt{K_{1}} r\right) K_{1} > 0 \eqno(13b)$$
$$f_{0}(r) = cosh^{2}\left(\sqrt{K_{1} r}\right), f_{1}(r) =
sinh^{2}\left(\sqrt{K_{1} r}\right) K_{1} < 0 \eqno(13c)$$
and assume that the curvature is small. Why we assume this will
be evident shortly. The equations $(12)$ imply
$$H^{2}_{01r} = \frac{K(D - 1)}{2} f_{0} f_{1} \eqno(14)$$
$$and \enskip\enskip\partial^{2}_{r} \Phi = 0\enskip i.e.\enskip \Phi = \alpha r + \beta \eqno(15)$$
$(9b)$ then yields $^{6}$
$$\alpha^{2} = K(1 - D)D + \left({1\over 4}\right) K(1 - D) \eqno(16)$$
This means that $\alpha^{2}$ is of the order of $K$. The equation
of motion for $H_{\mu\nu\lambda}$ (i.e.$(9c)$) is satisfied for
the derived solutions $(14)$ and $(15)$ except when $\mu = 1, \nu =
0$ and this non-vanishing part is
$$\alpha \left[\frac{K(D - 1)}{2} f_{0} f_{1}\right]^{1\over 2} \eqno(17)$$

For $B_{\mu\nu} \not= 0$, these solutions are valid and can be
made compatible with the equations of motion as follows. In the
light of $(13b,c), (16)$ and the assumption of small $K, (17)$ is of
the order of $K^{5\over 2}$. Retaining upto terms linear in the
curvature, $(17)$ may be ignored and the equations of motion
satisfied. So in this approximation of small curvature, we can
have maximally symmetric solutions with $f_{0}$ and $f_{1}$, in
conjunction with $\Phi = \alpha r + \beta$ and a non-vanishing
$B_{\mu\nu}$. We confine ourselves to the case of positive
curvature. Identical conclusions also hold for negative
curvature. $(13a)$ and $(12b)$ give the solution for $B_{\mu\nu}$ as $^{6}$
$$\hat{B}_{01} = - \left({1\over 4}\right) cos \left[2
\sqrt{K_{1}} r\right] \eqno(18)$$
so our starting solution with a maximally symmetric metric is
$(13), (15)$ and $(18)$. Using $(8)$, the twisted solutions are
$$\hat{G'} = \left(\matrix{
-F_{0}	&0\cr
0	&F_{1}
\cr}\right) \eqno(19a)$$

$$F_{0} = \frac{f_{0}}{\left[1 + \left(1 - f_{0}f_{1}\right) sinh^{2}\theta +
\hat{B}_{01} \left(\hat{B}_{01} sinh^{2}\theta + sinh 2\theta\right)\right]}$$
$$F_{1} = \frac{f_{1}}{\left[1 + \left(1 - f_{0}f_{1}\right) sinh^{2}\theta +
\hat{B}_{01} \left(\hat{B}_{01} sinh^{2}\theta + sinh 2\theta\right)\right]}$$
$$\hat{B'} = \frac{\left({1\over 2}\right)\left(1 - f_{0}f_{1} +
\hat{B}^{2}_{01}\right) sinh 2\theta + \hat{B}_{01} cosh
2\theta}{1 + \left(1 - f_{0}f_{1}\right) sinh^{2}\theta +
\hat{B}_{01} \left(\hat{B}_{01} sinh^{2}\theta + sinh
2\theta\right)} \left(\matrix{0		&1\cr	-1	&0
\cr}\right) \eqno(19b)$$
$$\Phi' = \alpha r + \beta - ln \left[1 + \left(1 -
f_{0}f_{1}\right) sinh^{2}\theta +
\hat{B}_{01}\left(\hat{B}_{01} sinh^{2}\theta + sinh
2\theta\right)\right] \eqno(19c)$$
The analogues of (12a) is now
$$H'^{\enskip 2}_{01r} = K_{1}' \frac{(D - 1)}{2} F_{0} F_{1} \eqno(20)$$
and this may be solved to get the antisymmetric tensor field as :
$$\hat{B'}_{01} = -
\left[\frac{2\left[\frac{K'_{1}}{K_{1}}\right]^{1\over
2}}{\left[sinh\theta \left(12 + 11 sinh^{2}\theta\right)^{1\over
2}\right]}\right]$$
$$tan^{-1}\left[\frac{\left(5 sinh\theta
cos2\sqrt{K_{1}} r - 4 cosh \theta\right)}{2\left(12 + 11
sinh^{2}\theta\right)^{1\over 2}}\right] \eqno(21)$$
(19b) and (21) can never be matched to be identical for any
value of $\theta$. So maximal symmetry is not preserved under
the 0(d) $\otimes$ 0(d) transformation.

\centerline {\bf IV.The Meaning of Maximal Symmetry when torsion is present}

We now give a generalisation of the meaning of maximal symmetry in presence of
torsion $^{7}$.In $N$ dimensions, a metric that admits the maximum number $N(N + 1)/2$ of
Killing vectors is said to be maximally symmetric. A maximally symmetric space
is homogeneous and isotropic about all points $^{1)}$.Such spaces are
of natural interest in the general theory of relativity as they correspond to
spaces of globally constant curvature which in turn is related to the concepts
of homogeneity and isotropy. The requirement of isotropy and homogeneity leads
to maximally symmetric metrics in the context of standard cosmologies - the
most well known being the Robertson-Walker cosmology. However, in the presence
of torsion there is a drastic change in the scenario and one needs to {\it redefine}
maximal symmetry itself. Here we do this in a way such that {\it the usual physical
meaning of maximal symmetry remains the same.} The only requirement is that the
torsion fields satisfy some mutually consistent constraints.We shall also give
examples of toy models where these ideas can be realised by determining the
torsion fields, the metric and the Killing vectors.

The possible implications of torsion have been discussed extensively by
G.Esposito $^{8}$. He studied a model of gravity (cast in hamiltonian form)
with torsion in a closed Friedmann-Robertson-Walker universe and obtained the
full field equations. He showed that the torsion leads to a primary constraint
linear in the momenta and a secondary constraint quadratic in the momenta (1989).
Subsequently, using the generalised Raychaudhuri equation he showed that Hawking's
Singularity Theorem can be generalised if the torsion satisfies some conditions (1990).
Esposito also studied the geometry of complex spacetimes with torsion (1993).
Alternative approaches involving torsion have been discussed by F.Hehl $^{9}$.

Presence of torsion implies that the affine connections
$\bar{\Gamma}^{\alpha}_{\mu\nu}$ are asymmetric and contain an
antisymmetric part $H^{\alpha}_{\mu\nu}$ in addition to the
symmetric term $\Gamma^{\alpha}_{\mu\nu}$  $^{8, 12}$ :
$$\bar{\Gamma}^{\alpha}_{\mu\nu} = \Gamma^{\alpha}_{\mu\nu} +
H^{\alpha}_{\mu\nu} \eqno(22)$$
$H_{\alpha\mu\nu} = \partial_{(\alpha}B_{\mu\nu)}$, the
background torsion, is completely arbitrary to start with.
$B_{\mu\nu}$ is the second rank antisymmetric tensor field.

Defining covariant derivatives with respect to
$\bar{\Gamma}^{\alpha}_{\mu\nu}$ we have for a vector field
$V_{\mu}$ :
$$V_{\mu;\nu;\beta} - V_{\mu;\beta;\nu} = -\bar{R}^{\lambda}_{\mu\nu\beta}
V_{\lambda} + 2 H^{\alpha}_{\beta\nu} V_{\mu;\alpha} \eqno(23)$$
$$\bar{R}^{\lambda}_{\mu\nu\beta} = R^{\lambda}_{\mu\nu\beta} +
\tilde{R}^{\lambda}_{\mu\nu\beta} \eqno(24a)$$
$$R^{\lambda}_{\mu\nu\beta} = \Gamma^{\lambda}_{\mu\nu, \beta} -
\Gamma^{\lambda}_{\mu\beta, \nu} +
\Gamma^{\alpha}_{\mu\nu}\Gamma^{\lambda}_{\alpha\beta} -
\Gamma^{\alpha}_{\mu\beta}\Gamma^{\lambda}_{\alpha\nu} \eqno(24b)$$
$$\tilde{R}^{\lambda}_{\mu\nu\beta} = H^{\lambda}_{\mu\nu, \beta} -
H^{\lambda}_{\mu\beta, \nu} +
H^{\alpha}_{\mu\nu}H^{\lambda}_{\alpha\beta} -
H^{\alpha}_{\mu\beta}H^{\lambda}_{\alpha\nu}$$
$$\phantom{\tilde{R}^{\lambda}_{\mu\nu\beta}} +
\Gamma^{\alpha}_{\mu\nu}H^{\lambda}_{\alpha\beta} -
H^{\alpha}_{\mu\beta}\Gamma^{\lambda}_{\alpha\nu} +
H^{\alpha}_{\mu\nu}\Gamma^{\lambda}_{\alpha\beta} -
\Gamma^{\alpha}_{\mu\beta}H^{\lambda}_{\alpha\nu} \eqno(24c) $$
The generalised curvature $\bar{R}^{\lambda}_{\mu\nu\beta}$ does
not have the usual symmetry (antisymmetry) properties. The last term on the
right hand side of $(23)$ is obviously a tensor. Hence $\bar{R}^{\lambda}_{\mu\nu\beta}$
is also a tensor.

Let $\xi_{\mu}$ be a Killing vector defined through the Killing condition:
$$\xi_{\mu\enskip ;\enskip\nu} + \xi_{\nu\enskip ;\enskip\mu} =
0\eqno(25)$$
This condition is preserved also in the presence of
$H^{\alpha}_{\mu\nu}$.

Equation $(23)$ for a Killing vector hence takes the form :
$$\xi_{\mu;\nu;\beta} - \xi_{\mu;\beta;\nu} = -
\bar{R}^{\lambda}_{\mu\nu\beta} \xi_{\lambda} +
2 H^{\alpha}_{\beta\nu}	 \xi_{\mu;\alpha}\eqno(26)$$
The $H^{\alpha}_{\beta\nu}$ are arbitrary to start with and so we may choose
them to be such that
$$H^{\alpha}_{\beta\nu}	 \xi_{\mu ;\alpha} = 0 \eqno(27a)$$
This is a constraint on the $H^{\alpha}_{\beta\nu}$ and not the
$\xi_{\nu}$ and is essential for the existence of maximal symmetry in
presence of torsion as we shall shortly see.

Therefore :
$$\xi_{\mu;\nu;\beta} - \xi_{\mu;\beta;\nu} = -
\bar{R}^{\lambda}_{\mu\nu\beta} \xi_{\lambda}\eqno(28)$$
We now impose the cyclic sum rule on $\bar{R}^{\lambda}_{\mu\nu\beta}$ :
$$\bar{R}^{\lambda}_{\mu\nu\beta} + \bar{R}^{\lambda}_{\nu\beta\mu} +
\bar{R}^{\lambda}_{\beta\mu\nu} = 0 \eqno(29a)$$
The constraint (29a) implies
$$\tilde{R}^{\lambda}_{\mu\nu\beta} + \tilde{R}^{\lambda}_{\nu\beta\mu} +
\tilde{R}^{\lambda}_{\beta\mu\nu} = 0 $$
i.e.
$$ H^{\lambda}_{\mu\nu,\beta} +
H^{\alpha}_{\mu\nu}\bar{\Gamma}^{\lambda}_{\alpha\beta} +
H^{\lambda}_{\nu\beta,\mu} +
H^{\alpha}_{\nu\beta}\bar{\Gamma}^{\lambda}_{\alpha\mu} +
H^{\alpha}_{\beta\mu,\nu} +
H^{\alpha}_{\beta\mu}\bar{\Gamma}^{\lambda}_{\alpha\nu} =
0\eqno(29b) $$
Adding $(28)$ and its two cyclic permutations and using (25),
$$\xi_{\mu;\nu;\beta} = -
\bar{R}^{\lambda}_{\beta\nu\mu}\xi_{\lambda}\eqno(30)$$
Then following usual arguments $^{4}$
$$\xi^{n}_{\mu}(x) = A^{\lambda}_{\mu}(x\enskip ;\enskip X)
\xi^{n}_{\lambda}(X) + C^{\lambda\nu}_{\mu} (x\enskip;\enskip X)
\xi^{n}_{\lambda ;\nu}(X)\eqno(31)$$
where $A^{\lambda}_{\mu}$ and $C^{\lambda\nu}_{\mu}$ are
functions that depend on the metric and torsion and $X$, but not
on the initial values $\xi_{\lambda}$ (X) and $\xi_{\lambda ;
\nu}$ (X), and hence are the same for all Killing vectors. Also
note that the torsion fields present in $A^{\lambda}_{\mu}(x ;X)$ and
$C^{\lambda\nu}_{\mu} (x ; X)$ obey the constraint $(29b)$. A
set of Killing vectors $\xi^{n}_{\mu}$ (x) is said to be
independent if they do not satisfy any relations of the form
$\sum_{n} d_{n} \xi^{n}_{\mu}(x) = 0$, with constant coefficients
$d_{n}$. It therefore follows that there can be at most ${N(N +
1)\over 2}$ independent Killing vectors in N dimensions, even in
the presence of torsion provided the torsion fields satisfy the
constraints $(27a)$ and $(29b)$	 $^{7}$.

Consider the constraints $(27a)$ and $(29b)$. $(27a)$ ensures that the generalised
curvature $\bar{R}^{\lambda}_{\mu\nu\beta}$ behaves in the same way as
$R^{\lambda}_{\mu\nu\beta}$ so far as the behaviour of the quantity
$V_{\mu;\nu;\beta} - V_{\mu;\beta;\nu}$ is concerned. We shall soon see that
this constraint is essential for the existence of maximal symmetry in presence
of torsion. $(29b)$ follows by demanding the cyclicity property of
$\bar{R}^{\lambda}_{\mu\nu\beta}$  which again is the usual property of a curvature
tensor $R^{\lambda}_{\mu\nu\beta}$. {\it These are the only two constraints
necessary to have maximal symmetry in presence of torsion and follow from
usual properties that a curvature tensor is  expected to possess. All this
therefore implies that the constraints $(27a)$ and $(29b)$ are unique.}

$\bar{R}^{\lambda}_{\mu\nu\beta}$ is antisymmetric in the indices
$\nu$ and $\beta$ once the above constraints are satisfied.
Proceeding as in ref.[4] we have :

$$(N-1) \bar{R}_{\lambda\mu\nu\beta} = \bar{R}_{\beta\mu}g_{\lambda\nu} -
\bar{R}_{\nu\mu}g_{\lambda\beta}$$
i.e.$$(N-1) R_{\lambda\mu\nu\beta} + (N-1) \tilde{R}_{\lambda\mu\nu\beta}$$
$$= R_{\beta\mu} g_{\lambda\nu} - R_{\nu\mu} g_{\lambda\beta} +
\tilde{R}_{\beta\mu} g_{\lambda\nu} -
\tilde{R}_{\nu\mu} g_{\lambda\beta}\eqno(32)$$
$R_{\lambda\mu\nu\beta}, R_{\beta\mu}$ are functions of the
symmetric affine coefficients $\Gamma$ only, whereas
$\tilde{R}_{\lambda\mu\nu\beta}$, $\tilde{R}_{\beta\mu}$ are
functions of both $\Gamma$ and $H$. Moreover, to start with $\Gamma$ and $H$
are independent. Broadly, the solution space of equation $(32)$ consists of
(a) solutions with $H$ determined by $\Gamma$ or vice versa (b) solutions where
$H$ and $\Gamma$ are independent of each other. All these solutions lead to
maximally symmetric spaces even in the presence of torsion.

We shall now illustrate that the smaller subspace (b) of these solutions
enables one {\it to cast the definition of maximal symmetry in the presence of
torsion in an exactly analogous way to that in the absence of torsion.}
A particular set of such solutions of  $(32)$ can be obtained
by  equating corresponding terms on both sides to get:
$$(N-1) R_{\lambda\mu\nu\beta} = R_{\beta\mu} g_{\lambda\nu} -
R_{\nu\mu} g_{\lambda\beta}\eqno(33a)$$
$$(N-1) \tilde{R}_{\lambda\mu\nu\beta} = \tilde{R}_{\beta\mu} g_{\lambda\nu} -
\tilde{R}_{\nu\mu} g_{\lambda\beta}\eqno(33b)$$
The above two equations lead to
$$R_{\lambda\mu\nu\beta} = \frac{R^{\alpha}_{\alpha}\left(g_{\lambda\nu} g_{\mu\beta} -
g_{\lambda\beta} g_{\nu\mu}\right)}{N(N - 1)}\eqno(34a)$$
$$\tilde{R}_{\lambda\mu\nu\beta} = \frac{\tilde{R}^{\alpha}_{\alpha}
\left(g_{\lambda\nu} g_{\mu\beta} - g_{\lambda\beta} g_{\nu\mu}\right)}{N(N
-1)}\eqno(34b)$$
Note that
$$\tilde{R}^{\alpha}_{\alpha\nu\beta} = 0 \eqno(35a)$$
$$\tilde{R}^{\alpha}_{\mu\nu\alpha} = H^{\alpha}_{\mu\nu,\alpha} -
H^{\alpha}_{\mu\gamma} H^{\gamma}_{\alpha\nu} +
H^{\alpha}_{\mu\nu} \Gamma^{\gamma}_{\alpha\gamma} -
H^{\alpha}_{\gamma\nu} \Gamma^{\gamma}_{\mu\alpha} -
H^{\alpha}_{\mu\gamma} \Gamma^{\gamma}_{\alpha\nu} \eqno(35b)$$
Consider the constraint $(29b)$ with $\beta=\lambda$.This gives :
$$H^{\alpha}_{\mu\nu,\alpha} +
H^{\alpha}_{\mu\nu} \Gamma^{\gamma}_{\alpha\gamma} +
H^{\alpha}_{\nu\gamma} \Gamma^{\gamma}_{\alpha\mu} -
H^{\alpha}_{\mu\gamma}	\Gamma^{\gamma}_{\alpha\nu} = 0\eqno(36)$$
It is straightforward to verify that this implies
$\tilde{R}_{\mu\nu} = \tilde{R}_{\nu\mu}$. Hence $(29b)$ implies that
$\tilde{R}_{\mu\nu}$ is symmetric.  Under these circumstances
$$R_{\mu\nu} = \left({1\over N}\right) g_{\mu\nu}R^{\alpha}_{\alpha}\eqno(37a)$$
$$\tilde{R}_{\mu\nu} = \left({1\over N}\right) g_{\mu\nu}\tilde{R}^{\alpha}_{\alpha}
= H^{\alpha}_{\mu\beta} H^{\beta}_{\alpha\nu}\eqno(37b)$$
$$\tilde{R} = \tilde{R}^{\alpha}_{\alpha} =
g^{\mu\nu}\tilde{R}_{\mu\nu} = g^{\mu\nu} H^{\alpha}_{\mu\beta}
H^{\beta}_{\alpha\nu} = H^{\nu\alpha}_{\beta}
H^{\beta}_{\alpha\nu} \eqno(37c)$$

Now using arguments similar to those given in Ref.[4] for the
Bianchi identities we can conclude that $^{7}$
$$\bar{R}_{\lambda\mu\nu\beta} =
R_{\lambda\mu\nu\beta} + \tilde{R}_{\lambda\mu\nu\beta} =
\left(K+ \tilde{K}\right)\left(g_{\lambda\nu}g_{\mu\beta} - g_{\lambda\beta}g_{\mu\nu}\right) =
\bar{K}\left(g_{\lambda\nu}g_{\mu\beta} -
g_{\lambda\beta}g_{\mu\nu}\right)\eqno(38)$$
$$R^{\alpha}_{\alpha} = constant = K N (1 - N)\eqno(39a)$$
$$\tilde{R}^{\alpha}_{\alpha} = H^{\mu\lambda}_{\beta} H^{\beta}_{\lambda\mu}=
constant = \tilde{K} N (1 - N)\eqno(39b)$$
$$\bar{K} = K + \tilde{K} = constant\eqno(39c)$$
(In deriving the above results from  the Bianchi identities we have used
the fact that for a flat metric the curvature constant $K=0$.Hence demanding
$\bar{K}=0$ for a (globally) zero curvature space means that $\tilde{K}=0$
which in turn means that the torsion must vanish.)

We now discuss two simple toy models where the torsion field which satisfies
$(29b)$ and$(39b)$and is also consistent with $(27a)$. First note that {\it any }
non vanishing torsion is always consistent with $(27a)$ because
$$H^{\alpha}_{\beta\nu}	 \xi_{\mu ;\alpha} = 0\eqno(27a)$$
implies
$$H^{\alpha}_{\beta\nu}	 \xi_{\alpha ;\mu} = 0 \eqno(27b)$$
through the Killing condition $(25)$.Adding $(27a)$ and $(27b)$ gives
$$H^{\alpha}_{\beta\nu}(\xi_{\mu ;\alpha} + \xi_{\alpha ; \mu}) = 0 $$
Hence {\it any non-zero torsion} is consistent with $(27a,b)$.
The torsion is an antisymmetic third rank tensor obtained from a second rank
antisymmetric tensor $B_{\mu \nu}$ as follows:
$$H^{\alpha}_{\mu\nu}= \partial^{\alpha}B_{\mu \nu}
+ \partial_{\mu}B_{\nu}^{\alpha} + \partial_{\nu}B^{\alpha}_{\mu}$$
{\bf Toy Model 1.}\\
Consider dimension $D=3$ and a general form for the metric as
$$ds^2 = -f_{0}(r) dt^2 + f(r) dr^2 + f_{1}(r) (dx^1)^2 \eqno(40)$$
So the metric components are
$$g_{00}=-f_{0}(r), g_{rr}=f(r),  g_{11}= f_{1}(r)$$
i.e. the metric components are functions of $r$ only.Further assume that
all fields, including $B_{\mu \nu}$ , depend only on $r$.Then the torsion is
just $H^{r}_{01}$.

It is straightforward to verify that with the our chosen metric the only
non-zero components of $\Gamma^{\alpha}_{\mu \nu}$ are
$\Gamma^{r}_{00}$, $\Gamma^{r}_{11}$, $\Gamma^{r}_{rr}$,
$\Gamma^{0}_{0r}$ and  $\Gamma^{1}_{1r}$.

Now $(29b)$ with  $\lambda=\beta$ gives $(36)$ which in the case under
consideration reduces to  (using the antisymmetry of $H$)
$$H^{r}_{01,r} + H^{r}_{01} \Gamma^{r}_{rr} + H^{0}_{1r} \Gamma^{r}_{00}
- H^{1}_{0r} \Gamma^{r}_{11} = 0 $$

We may write
$$ H^{0}_{1r} = g^{00}g_{rr} H^{r}_{01},
\enskip	 H^{1}_{0r} =-g^{11}g_{rr} H^{r}_{01}$$
Therefore $(36)$ becomes
$$H^{r}_{01,r} + H^{r}_{01}\enskip [\Gamma^{r}_{rr}+g^{00}g_{rr}\Gamma^{r}_{00}
+g^{11}g_{rr}\Gamma^{r}_{11}] = 0 \eqno(41a)$$
Using the values
$$\Gamma^{r}_{rr}=\partial_{r} ln f^{1/2} ;
\enskip \Gamma^{r}_{00}= (1/2)(1/f)\partial_{r} f_{0} ;
\enskip \Gamma^{r}_{11}=-(1/2)(1/f)\partial_{r} f_{1} $$
leads to
$$H^{r}_{01,r} + H^{r}_{01}\partial_{r}[ln(f/(f_{0}f_{1}))^{1/2}]=0\eqno(41b)$$
whose solution is
$$H^{r}_{01} = [f_{0}f_{1})/f]^{1/2}\eqno(42)$$
(Note that the torsion can be taken proportional to the completely antisymmetric
$\epsilon$ tensor in three dimensions as follows: $H_{r01}=g_{rr}H^{r}_{01}$, and
so can be written as $H_{r01}=[ff_{0}f_{1}]^{1/2} \epsilon_{r01}$ and this can
be further integrated to give the "magnetic field" as
$B_{01}=\epsilon_{01}\int dr [ff_{0}f_{1}]^{1/2}$, etc.)

It is immediately verified that
$$H^{r}_{01}H^{01}_{r}= -1$$
(i.e. a constant) thereby satisfying the constraint $(39b)$.

If $\xi$ denotes a Killing vector then the Killing equations are :
$$\xi^{r}\partial_{r}f_{0} + 2 f_{0} \partial_{0}\xi^{0} = 0\eqno(43a)$$
$$\xi^{r}\partial_{r}f + 2 f \partial_{r}\xi^{r} = 0\eqno(43b)$$
$$\xi^{r}\partial_{r}f_{1} + 2 f_{1} \partial_{1}\xi^{1} = 0\eqno(43c)$$
$$f_{1}\partial_{0}\xi^{1} - f_{0} \partial_{1}\xi^{0} = 0\eqno(43d)$$
$$f\partial_{0}\xi^{r} -  f_{0} \partial_{r}\xi^{0} = 0\eqno(43e)$$
$$f_{1}\partial_{r}\xi^{1} +  f \partial_{1}\xi^{r} = 0\eqno(43f)$$
The most general solutions are :
$$f_{0}=exp[-2\alpha \int dr f^{1/2}]\eqno(44a)$$
$$f_{1}=exp[-2\gamma \int dr f^{1/2}]\eqno(44b)$$
$$\xi^{r}=1/f^{1/2}\eqno(44c)$$
$$\xi^{0}=\alpha t + \eta(x^{1}) +\beta\eqno(44d)$$
$$\xi^{1}=\gamma x^{1} + \psi(t) +\delta\eqno(44e)$$
One set of  solution of these equations are:
$$f_{0}=f_{1}=constant\enskip  (i.e.\gamma=\alpha=0; \beta=-\delta)$$
so that $\xi^{0}$ and $\xi^{1}$ differ upto a sign. It can be readily verified
that the finite non-zero torsion given by $(42)$ satisfies all the constraints.

{\bf Toy Model 2.}\\
Suppose the space is flat in the symmetric part of the connection ,i.e.
$\Gamma^{\lambda}_{\mu\nu}=0$. Then $\bar{\Gamma}^{\lambda}_{\mu\nu}=
H^{\lambda}_{\mu\nu}$. The metric coefficients depend on the symmetric part
of the connection only and hence the metric may be taken as:
$$ds^2 = - dt^2 +  dr^2 + (dx^1)^2 \eqno(45)$$
The Killing equations now are:
$$\partial_{0}\xi^{0} = 0 ;
\partial_{r}\xi^{r} = 0 ;
\partial_{1}\xi^{1} = 0 ;
\partial_{0}\xi^{1} = \partial_{1}\xi^{0} ;
\partial_{0}\xi^{r} = \partial_{r}\xi^{0} ;
\partial_{r}\xi^{1} = -\partial_{1}\xi^{r} $$
One set of solutions for the Killing vectors consist of constant  vectors
with the components $\xi^{0}$ and $\xi^{1}$ differing upto a sign. With this
set, again a constant value for the torsion is consistent and satisfies all
the constraints.

The motive of these illustrations is to show that one can construct scenarios
with a finite non-zero value for the torsion and still have maximal symmetry.

Therefore, in the presence of torsion the criteria of maximal symmetry has been
generalised through the equations $(34b), (38)$ and $(39)$. The physical meaning is
still that of a globally constant curvature (which now also has a contribution
from the torsion).We emphasize that in $(39c)$ $K$ and $\tilde{K}$ {\it are
separately constants.} The torsion fields are subject to the constraints $(27),
(29b)$, and $(39b)$.We mention that relations exactly similar to eqs.$(37)$ have
been obtained in ref.[13] in a totally different context.

\centerline{\bf V.The Question of Duality Invariance In Presence Of Torsion}

Let us investigate whether this generalised maximal symmetry can be preserved
by the 0(d) $\otimes$ 0(d) twist. For simplicity we take D = 3. It can be shown
that for the metric of the type (13a) this is not possible. However, consider a
more general form viz.
$$ds^{2} = - f_{0}(r) dt^{2} + f_{1}(r) \left(dx_{1}\right)^{2} + f(r) dr^{2}\eqno(46)$$

For D = 3, the torsion $H^{\alpha}_{\mu\nu}$ is just $H^{ r}_{
01}$. It is readily verified that such a torsion is consistent
with (27) and satisfies (29b). (39b) now means
$$H ^{10} _{r} H ^{r} _{01} = constant = \vartheta \eqno(47)$$
Now, generalised maximal symmetry implies that
$$\bar{R}_{00} = (1 - N)\bar{K} G_{00} \eqno(48a)$$
$$\bar{R}_{11} = (1 - N)\bar{K} G_{11} \eqno(48b)$$
$$\bar{R}_{rr} = (1 - N)\bar{K} G_{rr} \eqno(48c)$$

These equations lead to respectively
$$\partial_{r}P - \partial_{r}Q	 + P^{2} - Q^{2} + RQ - RP = 0 \eqno(49a)$$
$$\partial_{r}Q + Q^{2} - QP - QR = 0 \eqno(49b)$$
$$\partial_{r}P + P^{2} - PQ - PR = 0 \eqno(49c)$$
where $P = \left({1\over 2}\right)\partial_{r}$ ln $f_{t} ; Q = \left({1\over
2}\right)\partial_{r}$ ln $f_{1} ; R = \left({1\over 2}\right)\partial_{r}$ ln $f$

The solutions to eqs.$(49a-49c)$ can be formally written as :
$$P = (1/2) [ A_{1}exp\{{-\int dr (P+Q-R)}\}+A_{2}exp\{{-\int dr(P-Q-R)}\}]\eqno(50a)$$
$$Q = (1/2) [A_{2}exp\{{-\int dr (P-Q-R)}\}-A_{1}exp\{{-\int dr (P+Q-R)}\}]\eqno(50b)$$
$$R = \partial_{r}ln (P-Q) + P + Q = \partial_{r}ln (P+Q) + P - Q \eqno(50c)$$
There can be many possible solutions to the above equations.We reject the
solution $Q=0$ i.e. $f_{1}=const.$ as it can be shown that the generalised
maximal symmetry cannot be preserved.

For  generalised maximal symmetry to prevail we have the
analogue of (12a) i.e. $\left(G_{00} = -f_{0}, G_{11} = f_{1}, G_{rr}
= f etc.\right)$
$$(H_{01r})^{2} = \frac{\bar{K}(1 - D)}{2}\enskip G_{00}\enskip G_{11}\enskip G_{rr}
\eqno(51a)$$
i.e.
$$\hat{B}_{01} = \left[\frac{\bar{K}(1 - D)}{2}\right]^{1\over 2} \int dr
\left[G_{00}\enskip G_{11}\enskip G_{rr}\right]^{1\over 2} \eqno(51b)$$
The beautiful thing is that the relations (51) are exactly
equivalent to (39b) under the identification of $\bar{K}$ with
$\tilde{K}$. Actually it can be readily shown that $K$ and
$\tilde{K}$ are proportional so that this identification is
consistent.

Now, under the O(d) $\otimes$ O(d) twist, the new fields are
$$(H'_{01r})^{2} = \frac{\bar{K'}(1 - D)}{2}\enskip G'_{00}\enskip G'_{11}\enskip G_{rr}
\eqno(52a)$$
i.e.
$$\hat{B'}_{01} = \left[\frac{\bar{K'}(1 - D)}{2}\right]^{1\over 2} \int dr
  \left[G'_{00}\enskip G'_{11}\enskip G_{rr}\right]^{1\over 2} \eqno(52b)$$
Here $G'_{00} = -F_{0}$, $G'_{11} = F_{1}$, $G_{rr} = f$, $F_{0}, F_{1}$
are given by $(19b)$ and $G_{rr} = f$ remains unaffected by the
$O(d)$ twist. This gives the freedom of chosing the function $f(r)$ to be such
that the equations $(19b)$, and $(49 - 52)$ are satisfied. Hence under these
circumstances the generalised maximal symmetry may  be preserved.

\centerline {\bf VI.Conclusion}

Maximally symmetric cosmological solutions have obvious physical implications
for the observable universe. Such solutions in the context of low energy string
effective theories already exist in the literature $^{10, 11}$. Tseytlin$^{10}$
studied time dependent solutions of the leading order string effective equations
for a non-zero central charge deficit and curved maximally space. Bento {\it et al}
$^{11}$ considered the effect of higher-curvature terms in the string low
energy effective actions on the maximally symmetric cosmological solutions of
the theory for various types of string theories.

Here we	 have shown that the $O(d) \otimes O(d)$ twist on maximally symmetric vacuum
solutions gives inequivalent solutions that are not maximally symmetric.
However, a generalised definition of maximal symmetry can be given (taking
into account torsion) and this may be preserved by the $O(d) \otimes O(d)$
transformation under certain conditions.

Our work is therefore crucially important in the context of string related
cosmology.We have seen that the background spacetime of string theory necessarily
has torsion in the form of a second rank antisymmetric tensor field.Even
if one starts with a torsion free background,duality transformation
automatically generates solutions with torsion.The concept of isotropic
and homogenous universe in cosmology which is realized usually in the form
of Robertson-Walker metric must therefore be changed when torsion is present.Our
work is thus a step towards a consistent cosmological background in the
context of string theory. The  ideas and formalism introduced in this work
requires further extension into various other aspects of observational
cosmology. Such work is under progress.

We thank the referee for references and extremely constructive suggestions-
in particular, the problem discussed in the toy model 2.

\end{document}